\documentclass[twocolumn,amsmath,amssymb,aps]{aastex62}
\usepackage[fleqn]{amsmath}
\usepackage{mathrsfs}

\usepackage{ctable}
\usepackage{graphicx}
\usepackage{dcolumn}
\usepackage{bm}
\usepackage{makecell}
\usepackage{footmisc}

\received{}
\revised{}
\accepted{}
\submitjournal{The Planetary Science Journal}

\shorttitle{The possible formation of Jupiter from supersolar gas}
\shortauthors{Aguichine et al.}

\begin{document}
	
\title{The possible formation of Jupiter from supersolar gas}

\author{Artyom Aguichine}
\email{artem.aguichine@lam.fr}
\affil{Aix Marseille Univ, Institut Origines, CNRS, CNES, LAM, Marseille, France}
\author{Olivier Mousis}
\affil{Aix Marseille Univ, CNRS, CNES, LAM, Marseille, France}

\author{Jonathan Lunine}
\affil{Department of Astronomy, Cornell University, Ithaca, NY 14853, USA}

\begin{abstract}
More than two decades ago, the Galileo probe performed \textit{in situ} measurements of the composition of Jupiter's atmosphere and found that the abundances of C, N, S, P, Ar, Kr and Xe were all enriched by factors of 1.5--5.4 times their protosolar value. Juno's measurements recently confirmed the supersolar N abundance and also found that the O abundance was enriched by a factor 1--5 compared to its protosolar value. Here, we aim at determining the radial and temporal evolution of the composition of gases and solids in the protosolar nebula (hereafter, PSN) to assess the possibility that Jupiter’s current composition was  acquired  via  the  direct  accretion of supersolar gases. To do so, we model the evolution of a 1D $\alpha-$viscous accretion disk that includes the radial transport of dust and ice particles and their vapors, with their sublimation and condensation rates, to compute the composition of the PSN. We find that the composition of Jupiter's envelope can be explained only from its accretion from PSN gas ($\alpha\le10^{-3}$), or from a mixture of vapors and solids ($\alpha>10^{-3}$). The composition of the PSN at 4 AU, namely between the locations of the H$_2$O and CO$_2$ icelines, reproduces the one measured in Jupiter between 100 and 300 kyr of disk evolution. Our results are found compatible with both the core accretion model, where Jupiter would acquire its metallicity by late accretion of volatile-rich planetesimals, and the gravitational collapse scenario, where the composition of proto-Jupiter would be similar to that of the PSN.


\end{abstract}


\section{Introduction} \label{sec:intro}

The Galileo probe performed \textit{in situ} measurements of the composition of Jupiter's atmosphere up to $\sim$22 bars of pressure \citep{Wo04}. These measurements indicated C, N, S, P, Ar, Kr and Xe abundances that were found 1.53 to 6.08 times higher than their protosolar values \citep{Mo18}, and subsolar abundances for Ne \citep[$\sim$0.10 protosolar;][]{Ma00} and O \citep[$\sim$0.46 protosolar;][]{Wo04}. The subsolar abundance of Ne in the upper troposphere has been explained by the sinking of liquid Ne with He \citep{Ma00}. It has also been proposed that the O subsolar abundance measured by the Galileo probe is indicative of a tar-- or a carbon--rich planet \citep{Lo04,Mo12}. However, the Galileo O measurement has also been attributed to the specificity of the probed region, which appeared water depleted \citep{Or98}. This hypothesis is supported by the recent H$_2$O abundance measurements by the microwave radiometer (MWR) aboard the Juno orbiter, which were performed in the equatorial region of Jupiter up to 30 bars of pressure \citep{Li20}. With this instrument, the H$_2$O and NH$_3$ abundances were found to be 1--5.1 and 2.6--2.9 times higher than the protosolar abundances of oxygen and nitrogen, respectively, at a 1-$\sigma$ level.
  
Two main models of giant planets formation currently exist, namely the core accretion model and the gravitational instability model. In the gravitational instability model, planetary formation begins with the protosolar nebula (PSN), breaking up due to its own self-gravity into giant protoplanets of solar composition, which then accrete planetesimals from the PSN \citep{Ca78,Bo97,He06}. In this scenario, Jupiter accretes its volatile content in 10-100 kyr. In the core accretion model, a solid core of $\sim$10 $M_\Earth$ is first formed from planetesimals, which then accretes gas and eventualy leftover planetesimals from the PSN \citep{Po96,Hu05}. Here, the accretion of planetesimals occurs over several Myr. Both scenarios rely on the accretion of planetesimals, which differs by the time and rate at which this accretion occurs. The amount and the nature of accreted material impact the final composition of the giant planet, and depend on the local composition of the PSN at the time of planet formation. Thus, measurements of the composition of gas giants from both our own solar system and other planetary systems, give constraints on the formation mechanism of those planets. More recently, a variant of the core accretion model has been proposed, in which the solid core is agglomerated from pebbles via streaming instability \citep{La12,Bo17,Jo17}. These studies focus on the final composition of giant planets formed in the framework of the pebble accretion scenario, using multiple prescriptions for accretion rates and planetary migration from 2D and 3D simulations \citep{Jo17,Bo17,Sc21a}. In this kind of model, the evolution of volatiles in the forms of vapors, dust or pebbles in the PSN is determined by the locations of the various icelines. Models of radial transport of dust, pebbles and gases in the PSN show that the sublimation and condensation around icelines change the local composition of solids and gases in the PSN, the accretion of which, in turn, affecting the final composition of the formed planets \citep{De17,Mo19,Ag20,Mo20}. The final composition of the gas giant then depends on many factors: composition of the accreted solids and gases, structure of the ices in solids (pure condensate, amorphous ice or clathrate hydrate), dissolution of accreted planetesimals in the envelope, atmospheric escape of lighter constituents, erosion of the core and so on \citep{Ow99,Ga01,Mo19,Mo21}.

The present study aims at determining the radial and temporal evolution of the composition of gases and solids in the PSN to assess the possibility that Jupiter's current metallicity was acquired via the direct accretion of supersolar gases. To do so, we explore the evolution of the PSN metallicity that allows it to reproduce that observed in Jupiter. The PSN material considered includes volatiles both in vapor and solid form. These solids are more or less gas-coupled, and their sizes range from a few microns to those of pebbles. All this material is assumed to form Jupiter's growing envelope, regardless of any particular formation mechanism. Our study allows us to investigate the relative contributions of the different volatile reservoirs --gases or solids-- to Jupiter's current metallicity, assuming it did not evolve with time after formation was completed. We assume that the icy part of dust and pebbles corresponds to pure condensates formed along the different icelines. We compare the most up-to-date measurements of the composition of Jupiter's atmosphere with the composition of the PSN computed with our 1D $\alpha$-viscous disk model \citep{Ag20,Mo20}. Our model provides us with constraints on Jupiter's formation time and location in the PSN, and gives insights on the composition and dynamics of the PSN.



\section{Overview of the model} \label{sec:model}
The model used for this study is the one described in \cite{Ag20} and \cite{Mo20}. Our code mimics the evolution of a 1D $\alpha$-viscous disk of surface density $\Sigma_\mathrm{g}(r)$, which corresponds to the mass density integrated over the azimuthal coordinate. The disk is assumed to be both isothermal and in hydrostatic equilibrium in the vertical direction. The disk's surface density is obtained by integrating the following equation \citep{Ly74}:

\begin{equation}
    \frac{\partial \Sigma_{\mathrm{g}}}{\partial t} = \frac{3}{r} \frac{\partial}{\partial r} \left[ r^{1/2} \frac{\partial}{\partial r} \left( r^{1/2} \Sigma_{\mathrm{g}} \nu \right)\right]. \label{eq:ofmotion}
\end{equation}

\noindent In this relation, the viscosity $\nu$ is computed in the framework of the $\alpha$-formalism:

\begin{equation}
\nu=\alpha \frac{c_\mathrm{s}^2}{\Omega_K}.
\label{eq:visc}
\end{equation}

\noindent where $\alpha$ is the viscosity parameter, $c_\mathrm{s}$ is the isothermal sound speed and $\Omega_K$ is the Keplerian frequency. The radial midplane temperature profile $T(r)$ includes the contributions of viscous heating, constant background radiation term, and, depending on the considered case (see Section \ref{sec:ccls}), direct irradiation from the Sun \citep{Na94,Hu05}.

At the beginning of the computation, the disk is uniformly filled with the trace volatile species H$_2$O, CO, CO$_2$, CH$_3$OH, CH$_4$, N$_2$, NH$_3$, PH$_3$, and H$_2$S whose abundances are calculated as follows. All elemental abundances are set to their protosolar values \citep{Lo09}. Carbon is distributed as CO:CO$_2$:CH$_3$OH:CH$_4$ = 10:30:1.67:1, a ratio consistent with ROSINA observations of comet 67P/Churyumov-Gerasimenko \citep{Le15,La19}, and the leftover oxygen is used to make H$_2$O. All phosphorus is in the form of PH$_3$ while half of the sulfur is in the form of H$_2$S \citep{Pa05}. The remaining sulfur is in refractory form and it is assumed that it sank to the core of Jupiter during accretion. By doing so, refractory sulfur presumably did not contribute to the S abundance of the envelope. The possible dissolution of these sulfides in the envelope is discussed in Sec. \ref{sec:ccls}. Nitrogen is also distributed between N$_2$ and NH$_3$ in the PSN, assuming N$_2$:NH$_3$ = 9:1, based on Spitzer observations of cloud cores and protostars \citep{Ob11,Po14}. Refractory dust, which does not sublimate or recondense, is also considered in our trace species. The dust abundance fills the condition $Z_\mathrm{tot} = Z_\mathrm{ice} + Z_\mathrm{dust}$ = 0.0153 \citep{Lo09}, where $Z_\mathrm{tot}$, $Z_\mathrm{ice}$ and $Z_\mathrm{dust}$ are the total metallicity of the protosun, the metallicity of the ices, and the metallicity of dust, respectively. The corresponding abundances are summarized in Table \ref{tab:abun}. At $t=0$, the solid-to-gas ratio is equal to $Z_\mathrm{tot}$ beyond all icelines, and $Z_\mathrm{dust}$ within the H$_2$O iceline, which is the innermost iceline.\\

\begin{table}[!ht]
\centering
\caption{Initial abundances of the trace species considered in our model. Values of $Z_\mathrm{ice}$ and $Z_\mathrm{dust}$ correspond to mass fractions of the molecular cloud.}
\begin{tabular}{lc}
\hline 
\hline		
Trace species   & $(\mathrm{X}/\mathrm{H})$$_\odot$\\
\hline 
		H$_2$O          & $1.409\times 10^{-4}$ \\
		PH$_3$          & $3.184\times 10^{-7}$ \\
		CO           & $6.499\times 10^{-5}$ \\
		CO$_2$          & $1.950\times 10^{-4}$ \\
		CH$_4$          & $6.499\times 10^{-6}$ \\
		CH$_3$OH        & $1.085\times 10^{-5}$ \\
		NH$_3$          & $8.185\times 10^{-6}$ \\
		N$_2$           & $3.683\times 10^{-5}$ \\
		H$_2$S          & $8.165\times 10^{-6}$ \\
		Ar           & $3.573\times 10^{-6}$ \\
		Kr           & $2.153\times 10^{-9}$ \\
		Xe           & $2.104\times 10^{-10}$ \\ \hline
		$Z_\mathrm{ice}$      & $0.01066$ \\
		$Z_\mathrm{dust}$ & $0.00464$\\
\hline 
\end{tabular}
\label{tab:abun}
\end{table}

The surface densities of trace species are evolved with an advection-diffusion equation \citep{Bi12,De17,Ag20,Mo20}:

\begin{equation}
\frac{\partial \Sigma_\mathrm{X}}{\partial t}+\frac{1}{r} \frac{\partial}{\partial r}\left[r\left(\Sigma_\mathrm{X} v_\mathrm{X}-D_\mathrm{X} \Sigma_{\mathrm{g}} \frac{\partial}{\partial r}\left(\frac{\Sigma_\mathrm{X}}{\Sigma_{\mathrm{g}}}\right)\right)\right]+\dot{Q}_\mathrm{X}=0,
\end{equation}

\noindent with species X being either in the form of solids $\Sigma_\mathrm{X,s}(r)$ or vapors $\Sigma_\mathrm{X,v}(r)$.
In vapor phase, the radial velocity $v_\mathrm{X,v}$ and the diffusion coefficient $D_\mathrm{X,v}$ are taken equal to those of the H$_2$/He gas $v_\mathrm{g}$ and $\nu$, respectively. The dynamics of dust grains is based on the two-population algorithm developed by \cite{Bi12}. The radial velocity of dust grains resulting from gas drag and radial drift is given by \citep{Bi12,Ag20}:

\begin{equation}
v_{\mathrm{X,s}} = \frac{1}{1+ \mathrm{St}^2}v_{\mathrm{g}} + \frac{ 2\mathrm{St}}{1+ \mathrm{St}^2}  v_{\mathrm{drift}}, \label{dust_velocity}
\end{equation}

\noindent where St is the particle's Stokes number \citep[see][for details]{Ag20} and we assume $D_\mathrm{X,s} = \frac{1}{1+\mathrm{St}^2}\nu \simeq \nu$. In the regimes considered here, the Stokes number follows the same trend as the particle's representative size, which only depends on the local solid-to-gas ratio \citep{Bi12,Ag20}. Small particles for which $\mathrm{St}\ll 1$ are dragged by the PSN gas ($v_{\mathrm{X,s}} \simeq v_{\mathrm{g}}$ in Eq. (\ref{dust_velocity})), but grow fast. Larger grains ($\mathrm{St}\sim 1$) experience less gas drag, but are subject to strong inward drift. At each time and location, one dust grain is assumed to be a mixture of all solid species, weighted by the surface density of each of them. Sublimation and condensation rates $\dot{Q}_\mathrm{X}$ are computed as in \cite{Dr17}.

\begin{figure}[!ht]
\resizebox{\hsize}{!}{\includegraphics[angle=0,width=5cm]{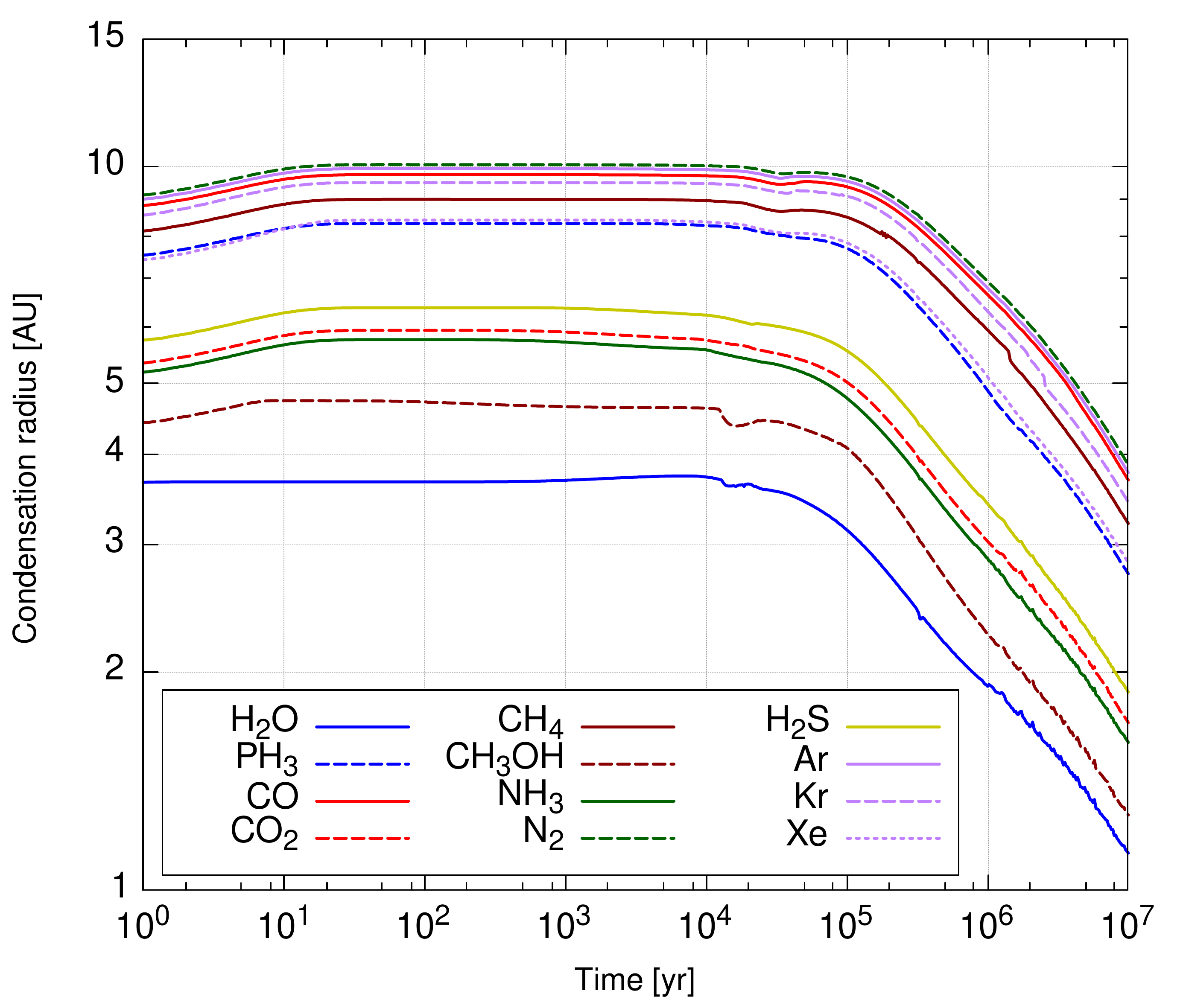}}
\caption{Positions of the icelines during the PSN evolution, assuming $\alpha=10^{-3}$. The relative positions of the icelines is not affected by the variation of the $\alpha$ value.}
\label{fig:icelines}
\end{figure}

We define the iceline of a given volatile species X as the heliocentric distance corresponding to the condition $\Sigma_\mathrm{X,s}=\Sigma_\mathrm{X,v}$\footnote{This condition defines the boundary between the vapor-- and solid--dominated regions.}. Figure \ref{fig:icelines} shows the time evolution of the positions of the icelines in the PSN. During the disk evolution, solid grains drift inward and sublimate, forming an iceline for each volatile species. As the disk cools down, the different icelines move inward in time. Due to the presence of a large number of small grains, the total icy surface available for sublimation or condensation is large and yields to significant sublimation and condensation rates. As a result, ices drifting within their icelines are almost instantaneously vaporized. Conversely, vapors diffusing beyond their icelines efficiently recondense into their solid forms. Consequently, each trace species is mostly in the solid phase beyond its iceline ($\Sigma_\mathrm{X,s}+\Sigma_\mathrm{X,v}\simeq\Sigma_\mathrm{X,s}$), and is predominantly in vapor phase within its iceline ($\Sigma_\mathrm{X,s}+\Sigma_\mathrm{X,v}\simeq\Sigma_\mathrm{X,v}$).

The disk evolution is ruled by its non-dimensional viscosity parameter $\alpha$, as its value impacts the disk size and dynamics. A higher value of $\alpha$ generates a more extended disk, and the growth in radial exchange of angular momentum also increases the values of the diffusion coefficients of the gas and trace species. The value of $\alpha$ does not directly impact the dynamics of solids. However, $\alpha$--values as low as those considered here imply a slow diffusion of the PSN gas. Consequently, a significant inward drift of large grains will result in a high dust--to--gas ratio in the innermost regions of the PSN, and a low dust--to--gas ratio in the outermost regions of the PSN. A larger viscosity also increases the viscous heating, pushing the different icelines slightly further. The values of $\alpha$ explored with our model is in the $\sim$10$^{-4}$--10$^{-2}$ range, based on models calibrated on disk observations \citep{Ha98,Hu05,De17}.

The supersolar metallicity of Jupiter's envelope is assumed to be acquired from the PSN as a mixture of H-He gas mixed with trace species in vapor and/or dust forms. To compare the PSN composition with that of Jupiter, the surface density profiles of the different species are converted into enrichment profiles. The time and radial evolution of the enrichment of a given species $X$ in the PSN is then defined by:

\begin{equation}
f_\mathrm{X,\phi}(r,t)  \equiv \frac{X/H(r, t)}{\left(X/H\right)_{(t = 0)}} = \frac{\Sigma_\mathrm{X,\phi} (r,t)/\Sigma_\mathrm{g}(r,t)}{\left(\Sigma_\mathrm{X}/\Sigma_\mathrm{g}\right)(t = 0)},
\label{eq:enrich_specie}
\end{equation}

\noindent where $r$, $t$, and $\phi$, are the heliocentric distance, time, and the phase of species $X$ (solid, vapor, or their sum), respectively. Species enrichments are then converted into elemental enrichments, which are normalized to their protosolar abundances. This normalization significantly decreases the sensitivity of our results regarding the choice of the initial composition.

In this study, we focus on the composition of the PSN at 4 AU, chosen as the formation location of Jupiter. This distance plausibly corresponds to a position in the disk beyond the H$_2$O iceline but inward of the icelines of all other trace species. In our model, this location is specifically between the icelines of H$_2$O and CO$_2$, which make up $\sim$52\% by mass of all trace species (volatiles + refractories) from Table \ref{tab:abun}. The dust-to-gas ratio in the vicinity of these icelines can easily become 2--3 times larger than the one derived from a mixture of protosolar composition (see Fig. \ref{fig:profiles}), easing the formation of proto-Jupiter's core via streaming instability \citep{Ya17}. Also, the local increase of the total gas surface density can generate hydrodynamic instabilities (vortices) that lead to gravitational instabilities by dust particles capture, forming giant protoplanets over timescales ($\sim$10 kyr) much shorter than the lifetime of the PSN \citep{Lo99,Lo14,Su15,Ba16}. Choosing a different radius (e.g. 5.2 AU, the current orbit of Jupiter) will not modify trends in our results as long as the chosen distance is between the icelines of H\textsubscript{2}O and CO\textsubscript{2}. This makes our model compatible with both the Nice model and the Grand Tack hypothesis, in which Jupiter is formed at 5.4 AU or 3.5 AU, respectively \citep{Br09,Wa11}.

\section{Results} \label{sec:results}

\begin{figure*}[!ht]
\resizebox{\hsize}{!}{\includegraphics[angle=0,width=5cm]{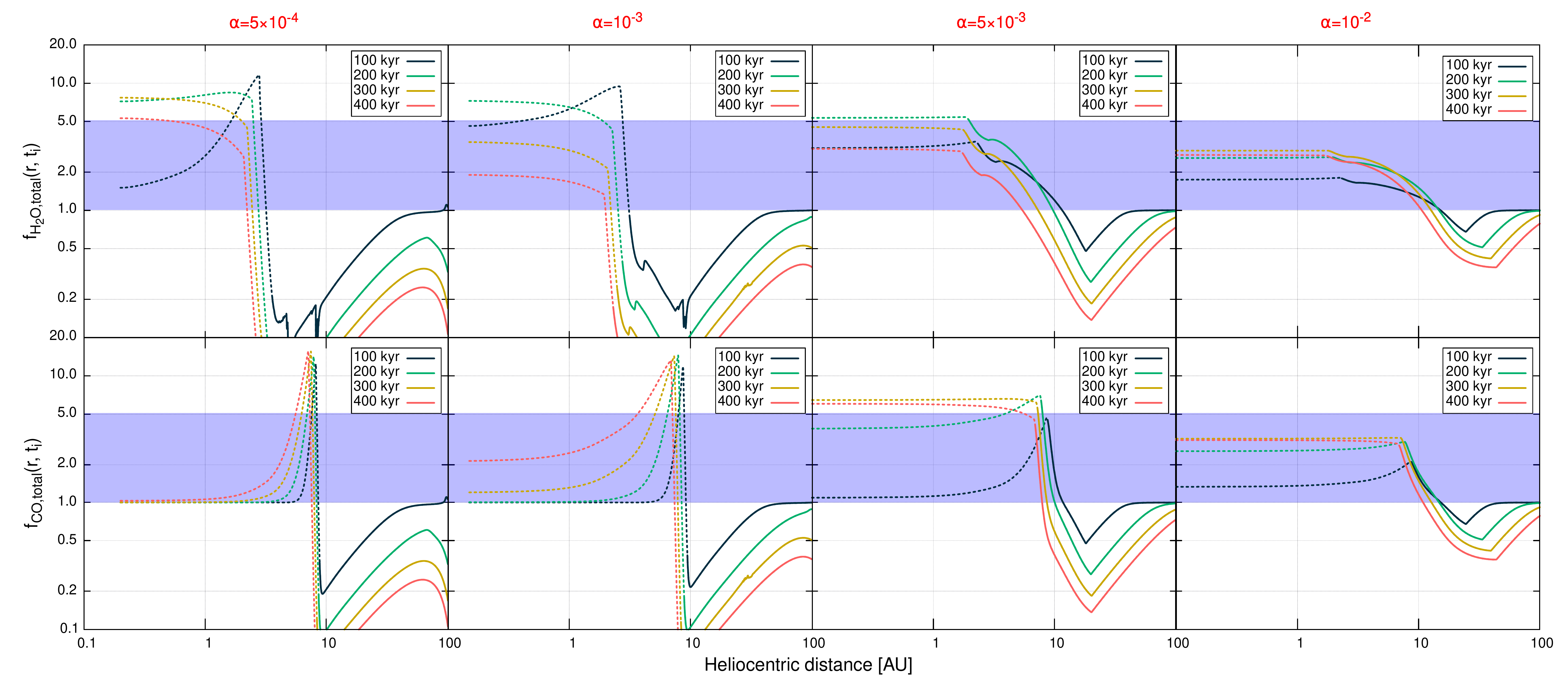}}
\caption{Top panels: water enrichment profiles $f_\mathrm{H_2O,\phi}(r,t)$ as a function of heliocentric distance at $t$ = 100, 200, 300, and 400 kyr of the PSN evolution, with $\phi$ = solid (solid lines) or vapor (dashed lines). Bottom panels: carbon monoxide enrichment profiles $f_\mathrm{CO,\phi}(r,t)$ as a function of heliocentric distance at the same epochs of the PSN evolution, with $\phi$ = solid (solid lines) or vapor (dashed lines). The blue bar corresponds to the H$_2$O abundance relative to the protosolar O abundance derived from Juno measurements at a 1-$\sigma$ level \citep{Li20}. Each column corresponds to a different value of $\alpha$.}
\label{fig:profiles}
\end{figure*}

Figure \ref{fig:profiles} represents the enrichment profiles of H$_2$O ($f_\mathrm{H_2O,\phi}(r,t)$) and CO ($f_\mathrm{CO,\phi}(r,t)$), both in solid and vapor phases, as a function of heliocentric distance at several epochs of the PSN evolution. The inward drift of solid grains leads to a significant decrease of the surface density of solids beyond the H$_2$O and CO icelines, and to an increase of the surface density of vapors near the icelines themselves. The formed vapors diffuse outward and recondense back in solid forms, creating an abundance peak at the iceline of each species. In our model, the diffusivity of vapors is taken to be equal to the diffusivity (viscosity) of the H$_2$+He gas, leading to a fast homogenization of the vapor concentrations in the inner regions of the disk. This region of nearly constant vapor to gas ratio is referred to as a enrichment plateau. A larger $\alpha$--value increases the efficiency of turbulent mixing, and generates more uniform surface density profiles of vapors in the inner disk. The enrichment plateau of H$_2$O vapor reaches a maximum after 100--400 kyr of the PSN evolution, depending on the $\alpha$--value. This plateau then decreases as vapors get accreted by the Sun. Similar trends can be observed for the CO enrichment profile. However, as vapors form at larger heliocentric distances, it takes more time for them to diffuse inward and become uniform. The blue shaded region shows Juno's measurements of H\textsubscript{2}O to H abundance, relative to the protosolar O/H abundance. This indicates the time and location at which Jupiter and the PSN have similar O content, if one considers that only H$_2$O or CO contributes to Jupiter's O abundance.

\begin{figure*}[!ht]
\resizebox{\hsize}{!}{\includegraphics[angle=0,width=5cm]{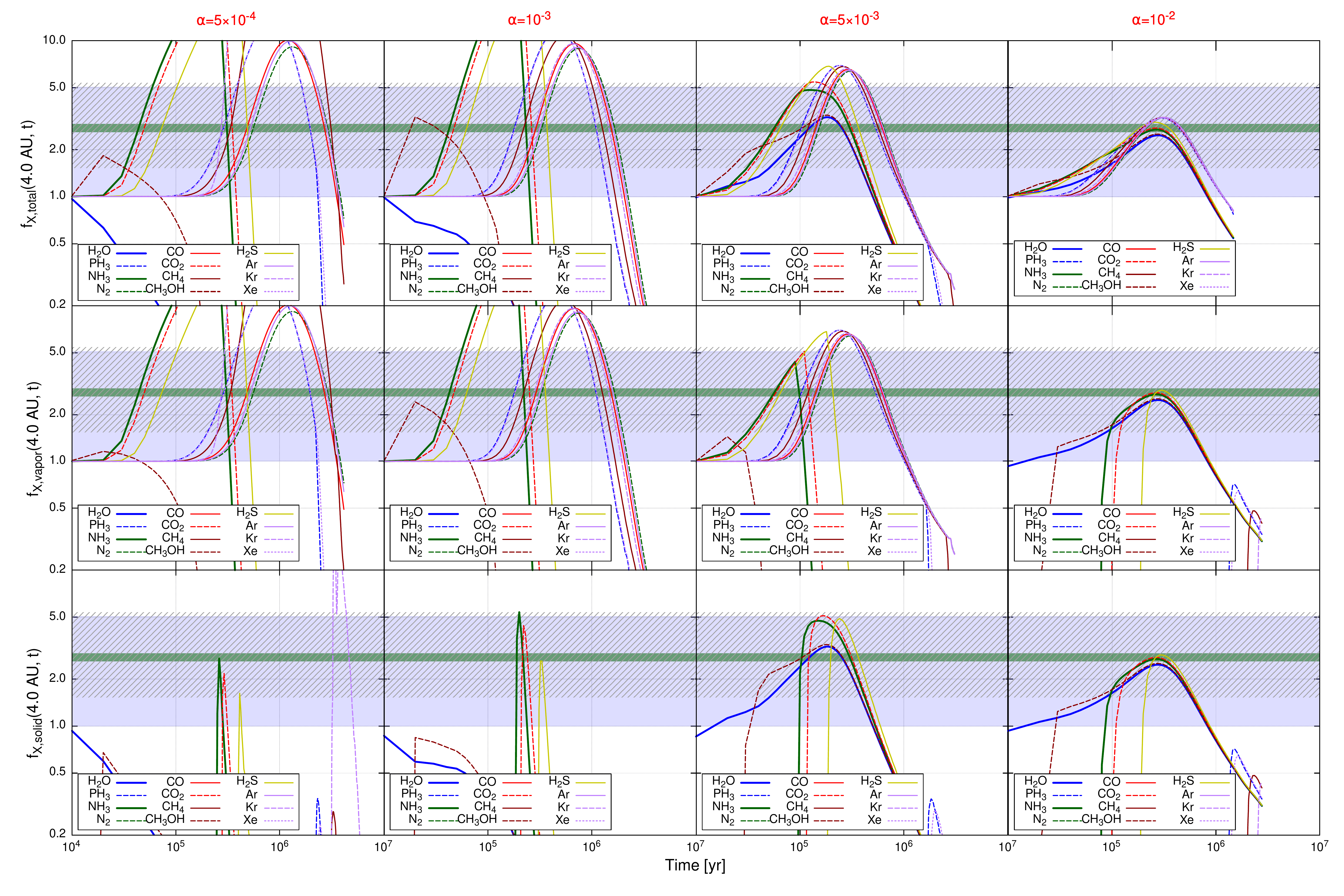}}
\caption{Time evolution of the trace species enrichments at the heliocentric distance of 4 AU. Top panels: species in solid+vapor phase $f_\mathrm{X,total}(4 {\rm AU}, t)$. Middle panels: species in vapor phase $f_\mathrm{X,vapor}(4 {\rm AU}, t)$. Bottom panels: species in solid phase $f_\mathrm{X,solid}(4 {\rm AU}, t)$. The blue and green bars correspond to the H$_2$O and NH$_3$ abundances relative to the protosolar O and N abundances, respectively, derived from Juno measurements at a 1-$\sigma$ level \citep{Li20}, respectively. The shaded bar corresponds to the range covered by S, C, P, Ar, Kr and Xe enrichments derived from spacecraft measurements \citep{At03,Mo18}. Each column corresponds to a different value of $\alpha$.}
\label{fig:profiles_species}
\end{figure*}

Figure \ref{fig:profiles_species} shows the time evolution of the enrichments ($f_\mathrm{X,\phi}(4 {\rm AU},t)$) of the different trace species $X$ listed in Table \ref{tab:abun} at the heliocentric distance of 4 AU. Top, middle and bottom panels correspond to the enrichments in solid+vapor phases, vapor phase, and solid phase, respectively. Here, the different volatile species embedded in drifting particles are released to the PSN gas phase when they cross their corresponding icelines, due to sublimation. As formed vapors diffuse inward, their abundances increase at 4 AU. Because the drift timescale of pebbles is much smaller than the diffusion timescale of the vapors, the vapor abundances (top panels of Figure \ref{fig:profiles_species}) only start to increase close to the positions of the various icelines (Fig. \ref{fig:icelines}). With time, the disk becomes depleted in the different species, implying the decrease of their abundances at 4 AU. If one of the icelines crosses the 4 AU distance, the corresponding species condenses, but the total abundance remains unchanged. Since the H$_2$O iceline is always located inside 4 AU, it is mostly present in solid form.

\begin{figure*}[!ht]
\resizebox{\hsize}{!}{\includegraphics[angle=0,width=5cm]{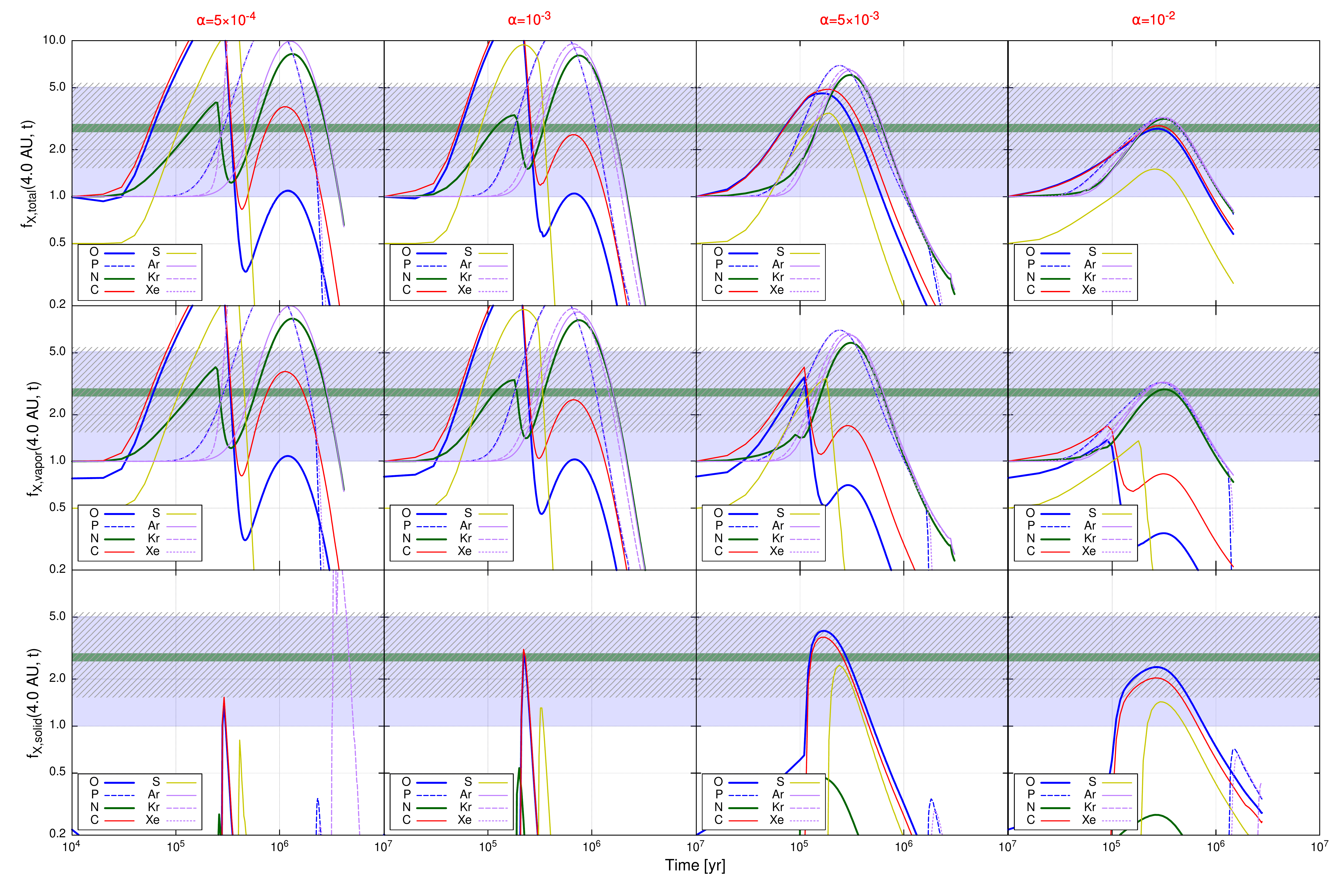}}
\caption{Same as Fig. \ref{fig:profiles_species}, but for elemental enrichments. When an element is assumed to be present in only one species (e.g. P in PH\textsubscript{3}, or noble gases), the enrichments are identical to those of Fig. \ref{fig:profiles_species}. Since C, N and O are distributed among several molecules, their enrichments exhibit several peaks. For example, when $\alpha \le 10^{-3}$, enrichment peaks of N in vapor+solid and vapor phases (top and middle rows) are due to NH\textsubscript{3} at 150 kyr, and N\textsubscript{2} at $\sim$1 Myr. When $\alpha \ge 5\times 10^{-3}$, the iceline of CO\textsubscript{2} moves below 4 AU after 100 kyr of PSN evolution, which results in a decrease of the C and O abundances in the vapor phase (middle row) and an increase in the C and O abundances in the solid phase (bottom row).}
\label{fig:profiles_elements}
\end{figure*}

Figure \ref{fig:profiles_elements} represents the time evolution of the elemental enrichments in solid+vapor phases (top panels), vapor phase (middle panels), and solid (bottom panels) at 4 AU, respectively. These enrichments are compared with the measured abundances of C, N, S, P and noble gases that are found to be between $\sim$1.5 and 5.4 times protosolar \citep{Mo18,Li20}. Due to its large uncertainty, the O enrichment is discussed in Section \ref{sec:ccls}. Since C, N and O are distributed among several molecules, their abundances present several enrichment peaks that are caused by the presence of multiple icelines.

In the solid+vapor phase, high $\alpha$--values produce similar elemental enrichments. When the $\alpha$--value is 10$^{-2}$, all elemental enrichments in our model match the 1.5--5.4 range at $t\sim 200$ kyr. The only exception to that rule is S, whose enrichment is too small, since half of this element is assumed to stay in refractory species. Because both C and O are enriched by the same factor, the C/O ratio remains protosolar ($\sim 0.5$). Similar trends are found when $\alpha = 5\times 10^{-3}$, but enrichment peaks are higher. Here, enrichments of P, N and noble gases match the 1.5--5.4 range at two distinct epochs located at the two sides of the peaks ($t$~=~100 and 400 kyr). Because of the fast accretion of CO$_2$ and H$_2$S onto the Sun, the PSN is depleted in C and S after 400 kyr of PSN evolution, but matches Jupiter's composition at $t=100$ kyr with supersolar O. When $\alpha$ is $\le 10^{-3}$, the effect of icelines becomes more important, and enrichment peaks are even higher. The C and O enrichment peaks are first caused by the enrichment peak of CO$_2$ at $t=150$ kyr (see Fig. \ref{fig:profiles_species}), and then caused by the enrichment peak of CO at $t\sim$2 Myr. Peaks in N enrichment are due to NH$_3$ at $t=150$ kyr and N$_2$ at $t\sim$1 Myr. Here, our results match the 1.5--5.4 enrichment range in a very narrow time domain around $t\simeq 300$ kyr.


When the $\alpha$--value is $\ge 5\times 10^{-3}$, C and O abundances in the vapor phase exhibit two peaks at 100 and 200 kyr. Because CO$_2$ is the main C-- and O--bearer, its dynamics greatly impacts the C and O abundances. At times earlier than 100 kyr, C and O abundances increase due to the inward diffusion of CO\textsubscript{2} vapor down to 4 AU. At later times, the CO\textsubscript{2} iceline moves below 4 AU, leading to a decrease of C and O abundances in the vapor phase. The second peak for C and O at 200 kyr is caused by the enrichment in CO vapor (by a factor of 10, as seen in Fig. \ref{fig:profiles_species}). However, due to the depletion in H$_2$O vapor, which is in solid phase at 4 AU, the C enrichment is twice higher than the O enrichment. This yields a C/O ratio of $\sim$1 (twice the protosolar ratio). Similarly to the behavior of CO$_2$, H$_2$S condenses at 150 kyr, and decreases the S abundance in the vapor phase around this location. When $\alpha$ is $\le 10^{-3}$, volatiles in the PSN are mostly in the form of vapor. The composition of the vapor phase is thus almost identical to the aforementioned  solid+vapor case, and our results reproduce the elemental enrichments measured in Jupiter at 300 kyr of the disk evolution at 4 AU.

Our calculations show that Jupiter's metallicity can be explained by the accretion of vapors only at $t\simeq 300$ kyr when the $\alpha$--value is lower or equal to $10^{-3}$. However, the dual accretion of vapors and pebbles at $t= 100--300$ kyr by the forming envelope is more likely when larger $\alpha$--values are considered. In those cases, solids may have contributed to the C and O enrichments in the forms of H$_2$O and CO$_2$ ices.



\section{Discussion and conclusion}  \label{sec:ccls}

In this work, we find that, depending on the choice of the value of the viscosity parameter, the metallicity of Jupiter's envelope can be explained only from its accretion from PSN gas ($\alpha$ $\le$ $10^{-3}$), or from a mixture of vapors and solids ($\alpha$ $>$ $10^{-3}$). The composition of the PSN at 4 AU reproduces Jupiter's metallicity at epochs in the 100--300 kyr range, regardless the $\alpha$--value. Specific times are summarized in Table \ref{tab:time}, depending on the $\alpha$-value and the phase of accreted volatiles. This enrichment is caused by an efficient transport of the volatile compounds from the outermost regions of the PSN to Jupiter's formation location.

\begin{table}
\centering
\caption{Time at which the composition of the PSN matches the metallicity of Jupiter's envelope, a function of $\alpha$ and the considered volatiles phase.}
\label{tab:time}
\begin{tabular}{|>{\centering\arraybackslash}p{1.5cm} | >{\centering\arraybackslash}p{2.7cm} | >{\centering\arraybackslash}p{2.7cm}|} \hline
$\alpha$-value & vapors only & vapors+solids \\ \hline\hline
$5\times 10^{-4}$ & $\sim$300 kyr & $\sim$300 kyr\tablenotemark{1} \\ \hline
$\phantom{{}5\times{}} 10^{-3}$ & $\sim$300 kyr & $\sim$300 kyr\tablenotemark{1} \\ \hline
$5\times 10^{-3}$ & $\varnothing$ & $\sim$100 or $\sim$300 kyr\\ \hline
$\phantom{{}5\times{}} 10^{-2}$ & $\varnothing$ & $\sim$200 kyr \\ \hline
\end{tabular}	
\tablenotetext{1}{Here, volatiles in the PSN are mostly in the form of vapors. Accretion of vapors or vapors+solids will result in almost identical metallicities of Jupiter's envelope.}
\end{table}

After 300 kyr of PSN evolution, elemental enrichments are either decreasing ($\alpha \ge 5 \times 10^{-3}$), or become uneven and never fit the 1.5--5.4 range together ($\alpha \le \times 10^{-3}$). In both cases, our results indicate that Jupiter's formation by accretion of PSN material later than $\sim$1.3 Myr is highly unlikely because after this epoch the PSN becomes depleted in volatiles. This result is consistent with the recent work of \cite{Sc21b} who proposed a formation model of Jupiter consistent with its current composition. These authors find that formation by pebble and gas accretion requires $\simeq 3$ Myr, resulting in planets with mostly subsolar abundances. To reproduce the observed metallicity, the authors emulated the accretion of $\sim$30 $M_\Earth$ of solids in an ad hoc way to form the current planet, a mass in agreement with the range found by \cite{Mo21}, assuming the solids were composed of several types of ices.

We outline that the concept of giant planet formation from pebbles still poses some issues. Such models consider Type I and Type II migration, which arise from the exchange of angular momentum between the planet and the viscous disk \citep{Kl12,Ba14}. This exchange of angular momentum creates a torque acting on the planet's orbit, resulting in a change of orbital parameters, e.g. the orbital radius. Both Type I and Type II migration share the same physical origin, but take place at different planet mass regimes, and mainly result in inward migration. Type I and Type II migration are expected to be significant with $\alpha$--values $\ge 5 \times 10^{-4}$. Numerical models show that a growing planetary embryo injected at 4 AU can migrate inward down to an orbit of $\sim$0.5 AU \citep{Bo17,Jo17,Sc21a}, impeding the formation of an envelope with a supersolar metallicity. These models show that with such small $\alpha$-values, stabilizing a Jupiter-like planet at the heliocentric distance of 5 AU requires that its core must have formed beyond 10 AU. However, this distance is located beyond the positions of all the icelines considered in our model, suggesting that forming a supersolar envelope from vapors or vapors + solids is incompatible with the radial transport of volatiles in the form of pebbles and vapors through multiple icelines. However, recent 2D and 3D simulations that model planet-disk interactions point out that Type I and II migration rates are several times lower than those given by prescriptions available in the literature \citep{Ch20,Le21,Ch22}. Those results indicate that an \textit{in situ} formation of Jupiter remains a viable hypothesis, which is the hypothesis adopted here. This highlights the current lack of understanding regarding the trajectory of planets formed in circumstellar disks. Alternative models, such as those requiring accretion of pebbles and planetesimals with appropriate compositions at different stages of the formation of Jupiter \citep{Al18,Mo21}, must be then considered to remain compatible with the core accretion model.

The local increase of the vapor surface densities of CO$_2$ and/or H$_2$O at 4 AU and $t=150$--$300$ kyr produces an axisymmetric bump of the gas surface density, which can trigger hydrodynamical instabilities leading to the formation of a vortex \citep{Lo99,Lo14}. Hydrodynamical simulations in protoplanetary disks show that vortices are efficient at concentrating pebbles in just a few hundreds of orbits ($\sim$10 kyr) \citep{Su15,Ba16}. This in turn could lead to a gravitational collapse that formed the young Jupiter with a composition similar to that of the PSN. Although our model is compatible with the development of such instabilities, current simulations of vortices are only efficient in the formation of planetesimals or planetary embryos. It is yet to be shown that vortices can fully form gas giant protoplanets that efficiently accrete gas and vapors from the PSN. Our model also does not include accretion of the envelope over time, which can produce a gradient of metallicity in the planet interior. This can give additional constraints on the accretion timescale required to reproduce Jupiter's metallicity profile \citep{De19}.

For $\alpha$--values larger than $5\times 10^{-3}$, the recondensations of CO$_2$ and H$_2$O at their respective icelines also produce local bumps in the solid--to--gas ratio (see Fig. \ref{fig:profiles}) that may trigger streaming instabilities \citep{La12,Jo17}. At these locations, Jupiter would be formed with a C-- and O--rich core and a C-- and O--poor envelope, while keeping supersolar overall C and O abundances. In this context, core erosion could explain Jupiter's supersolar metallicity \citep{St82,St85,Gu04,So17}. This could also produce a metallicity gradient in Jupiter's interior. We also note that, in both cases of streaming and hydrodynamic instabilities, refractory species, including sulfur compounds, may be partially or totally dissolved in the envelope. However, quantifying this effect requires modeling that is beyond the scope of the study.

Our calculations have been made assuming that the O determination in Jupiter ($1-5.1$ times protosolar O) corresponds to the 1-$\sigma$ level. At a 2-$\sigma$ level, this value becomes 0.1--7.5 times protosolar O \citep{Li20}. This implies the possibility that the O abundance in Jupiter's envelope could be subsolar as well. One way to obtain a subsolar O abundance at 4 AU is to increase the H$_2$O abundance by decreasing that of CO$_2$. This is the case when one assumes that CO was the main C-bearer in the outer PSN, as predicted by some thermochemical models \citep{Le80}. By doing so, more solid H$_2$O and less CO$_2$ vapor are available at 4 AU. Because most of the O is in form of H$_2$O ice, formation by accretion of gas and vapors only will result in a subsolar abundance of O in Jupiter's envelope.

In the broader context of the formation of Saturn, Figures \ref{fig:icelines} and \ref{fig:profiles} show that the PSN also presents supersolar abundances in volatile elements at $\sim$10 AU. This is due to the presence of multiple icelines at this location. The accumulation of volatiles at this heliocentric distance due to the inward drift of icy particles could also give a natural explanation of the observed composition of Saturn. In the meantime, the same process leads to a rapid depletion in volatiles of the outermost regions of the PSN, implying that our model cannot explain the high metallicity of the ice giants. We attribute this limitation of our model to the fact that gas and ice giants are formed by different formation processes.

We have shown that the metallicity of Jupiter's envelope can be achieved in the PSN, with volatiles being in vapor phase only, or a mixture of solids and vapors. Other processes (late accretion of planetesimals, core erosion etc.) may explain Jupiter's metallicity, but their consideration is beyond the scope of this paper. Further developments in the modeling of Jupiter's deep interior will provide a better knowledge of its interior structure and composition, which will help discriminating between various formation scenarios.

\section*{Aknowledgements}
OM acknowledges support from CNES. JL’s work was performed under a Juno subcontract to Cornell University from the Southwest Research Institute. We thank the anonymous referees for their comments that improved the quality of this paper. The project leading to this publication has received funding from the Excellence Initiative of Aix-Marseille Universit\'e -- A*Midex, a French “Investissements d’Avenir programme” AMX-21-IET-018.


\end{document}